\documentclass[debug,overfull,nobm,figures]{epl}
\usepackage{amssymb}
\newcommand{\text}[1]{\mbox{\scriptsize{#1}}}
\title{Demixing in binary mixtures of hard hyperspheres
}
\author{S. B. Yuste\inst{1}\thanks{E-mail: \email{santos@unex.es}} \and A. 
Santos\inst{1}\thanks{E-mail: \email{andres@unex.es}} \and M. L\'{o}pez de 
Haro\inst{2}\thanks{Also Consultant at Programa de Simulaci\'on 
Molecular del Instituto Mexicano del Petr\'oleo; 
E-mail: \email{malopez@servidor.unam.mx}}}
\shortauthor{S. B. Yuste \etal}
\institute{
\inst{1} Departamento de F\'{\i}sica,
Universidad de Extremadura -
 E-06071 Badajoz,  Spain\\
 \inst{2}
Centro de Investigaci\'on en Energ\'{\i}a, UNAM -
Temixco, Morelos 62580, M\'{e}xico
}

\begin{document}
\maketitle

\begin{abstract}
The phase behavior of binary  fluid mixtures  of hard
hyperspheres in four and
five dimensions is investigated.  Spinodal instability is found by
using a recent  and accurate prescription for the equation
of
state of the mixture that requires the equation of state of the single
component
fluid as input. The role played by the dimensionality on the possible metastability of the
demixing transition  with respect to a
fluid-solid transition is discussed.  The binodal curves in the 
pressure--chemical potential
representation are seen to lie on a common line, independent of the size
ratio.
\end{abstract}

The phase diagram of the simplest model of a binary mixture, that of
additive hard spheres, has not been  completely determined up to now
and remains as an
interesting open problem. Ever since the solution of the
Ornstein-Zernike
equation with the Rogers-Young closure for such a mixture  given by
Biben
and Hansen\cite{BH91}, the until then widespread and unchallenged belief
that these mixtures cannot phase separate into two fluid phases (which
was
supported by the analytical results of the Percus-Yevick approximation
\cite{L64}) has been seriously questioned. In fact, a wealth of
papers
dealing with this and related demixing problems have recently appeared
in
the
literature
\cite{FL92,LS93,R94,CB98b,C96,CAR00},
but the issue is still far
from
being settled. One of the key aspects of this fluid-fluid transition in
hard-sphere mixtures is that, if it exists at all, it must be entropy
driven. In contrast,  in other mixtures such as molecular mixtures, the
transition is driven by an asymmetry between the like- and
unlike-particle
interactions, { i.e.} it is energy driven, and at least the qualitative
phase behavior is rather well characterized\cite{RS82}.

Entropy driven demixing was nicely shown by Frenkel and Louis to occur in
a
simple lattice model for a binary hard-core mixture \cite{FL92}.
Experimental work has provided evidence that results for (asymmetric in
size) hard-sphere mixtures approximate rather well the thermodynamic
properties of real colloidal systems,  so it has been suggested that a
likely physical mechanism behind the entropy driven demixing transition
is
osmotic depletion. This notion, first introduced to explain phase
separation
in polymer-colloid mixtures, is included in the scaled particle theory
of
Lekkerkerker and Stroobants\cite{LS93},  which predicts a fluid-fluid
spinodal
instability for highly asymmetric mixtures but
 whose domain of validity is not clear yet.
  The depletion picture is also present in other more recent studies
\cite{R94,CAR00}.

A different approach to the demixing problem relies on the use of
equations
of state (EOS) for the hard-sphere mixture. It is well known that the
popular Boubl\'{\i}k-Mansoori-Carnahan-Starling-Leland (BMCSL) EOS
of
state\cite{B70} does predict no demixing. {On the other hand,} Coussaert 
and
Baus\cite{CB98b}
have proposed an
alternative EOS with improved virial behavior  that
predicts a fluid-fluid transition but at such high pressures that it is
argued to be metastable with respect to a stable fluid-solid transition.
In
this  Letter we also take the EOS approach  for binary
mixtures of
hard
additive {\em hyperspheres\/}. By looking into the higher dimensionality, one can gain insight into any thermodynamic phenomenology that extends
to such dimensionality  \cite{FP99}. We will 
show that indeed these systems
clearly exhibit a spinodal instability behavior.

Let us consider an  $N$-component mixture of (additive) hard
spheres in $d$
dimensions. The total number density is $\rho $, the set of molar
fractions
is $\{x_{i}\}$, and the set of diameters is $\{\sigma _{i}\}$. The 
packing fraction is $\eta =\sum_{i=1}^{N}\eta
_{i}=v_{d}\rho \langle \sigma ^{d}\rangle $, where $\eta _{i}=$
$v_{d}\rho
_{i}\sigma _{i}^{d}$ is the partial  packing fraction due to
species
$i $, $\rho _{i}=\rho x_{i}$ is the partial number density corresponding
to
species $i$, $v_{d}=(\pi /4)^{d/2}/\Gamma (1+d/2)$ is the volume of a
$d$-dimensional sphere of unit diameter, and $\langle \sigma ^{n}\rangle
\equiv
\sum_{i=1}^{N}x_{i}\sigma _{i}^{n}$. In previous work \cite{SYH99} we
have proposed a {\em simple\/} EOS for the mixture,
$Z_{\text{m}}(\eta )$,  consistent with a given
EOS for a  single component
system,
$Z_{\text{s}}(\eta )$, where $Z=p/\rho k_{B}T$ is the
compressibility
factor, $p$ being the pressure, $T$ the absolute temperature and $k_{B}$
the
Boltzmann constant. Our EOS reads
\begin{equation}
Z_{\text{m}}(\eta )=1+\left[ Z_{\text{s}}(\eta )-1\right] 2^{1-d}\Delta _{0}+
\frac{\eta }{1-\eta }\left[1-\Delta _{0}+\frac{1}{2}\Delta _{1}\right]
,
\label{4.1}
\end{equation}
\begin{equation}
\Delta _{p}\equiv\frac{\langle \sigma ^{d+p-1}\rangle }{\langle \sigma
^{d}\rangle ^{2}}\sum_{n=p}^{d-1}
\left(
\begin{array}{c}
d+p-1\\
n
\end{array}
\right)
\langle
\sigma
^{n-p+1}\rangle \langle \sigma ^{d-n}\rangle .  \label{4.2}
\end{equation}
In the
one-dimensional case, eq.~(\ref{4.1})  yields the {\em exact\/} result
$Z_{\text{m}}(\eta
)=Z_{\text{s}}(\eta )$.
Further, for $d=2$ and $3$,  it proved to be very
satisfactory
when a reasonably accurate  $Z_{\text{s}}(\eta )$ was taken
 \cite{SYH99,CCHW00}. In all instances
examined with different $Z_{\text{s}}(\eta )$,
however, no demixing is found for the resulting
EOS for the mixture in these dimensionalities.

\begin{figure}
\onefigure[scale=1,clip=]{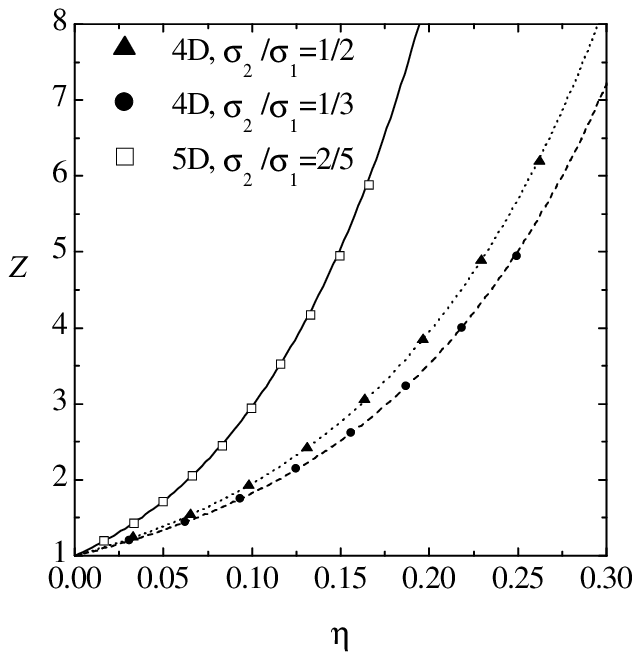}
\caption{Compressibility factor for  three equimolar mixtures
in 4D and 5D systems. Lines are the
theoretical results and symbols are
computer simulation data \protect\cite{Jose}.
\label{fig1}}
\end{figure}

 One-component fluids of hard hyperspheres ($d\geq 4$) have attracted
the attention of a number
of
researchers over the last twenty
years
\cite{FP99,LB82,LM90,MT84,CB86,FI81}.
Amongst other similarities, they share the property of hard disks and
spheres of exhibiting a first order freezing transition. This transition
occurs at a packing fraction $\eta _{\text{f}}$ that, relative to the
 close-packing fraction $\eta_{\text{cp}}$, decreases monotonically with
 increasing $d$. 
Perhaps the most accurate proposals to date for
$Z_{\text{s}}(\eta )$ in $d=4$ and $d=5$ are the semiempirical EOS
proposed by Luban and Michels \cite{LM90}:
\begin{equation}
Z_{\text{s}}(\eta )=1+b_{2}\eta \frac{1+\left[ b_{3}/b_{2}-\zeta
(\eta )b_{4}/b_{3}\right] \eta }{1-\zeta (\eta )(b_{4}/b_{3})\eta
+\left[
\zeta (\eta )-1\right] (b_{4}/b_{2})\eta ^{2}},  \label{4}
\end{equation}
where $b_{n}$ are (reduced) virial coefficients defined by the series
$Z_{\text{s}}(\eta )=1+\sum_{n=1}^{\infty }b_{n+1}\eta ^{n}$.
 Equation (\ref{4}) is consistent with the (known) exact first four virial
coefficients, regardless of the choice of $\zeta (\eta )$. Luban and
Michels
observed that the computer simulation data favor a {\em linear\/}
approximation for $\zeta (\eta )$ and by a least-squares fit procedure
found
$\zeta (\eta )=1.2973(59)-0.062(13)\eta/\eta_{\text{cp}}$ for $d=4$ and
$\zeta (\eta
)=1.074(16)+0.163(45)\eta/\eta_{\text{cp}}$ for $d=5$.
Substitution of eq.~(\ref{4}), using
the above linear fits for  $\zeta (\eta )$, into eq.~(\ref{4.1})
produces
the EOS of a
mixture of hard additive hyperspheres in $d=4$ and $d=5$ dimensions,
respectively. These EOS are the basis of our subsequent analysis.
As an illustration of the accuracy of our EOS, in Fig.\ \ref{fig1} we
display
the
comparison of the density dependence of $Z_{\text{m}}(\eta )$ as obtained in
this work
with the results of computer simulation\cite{Jose} for
 three equimolar  binary mixtures and different values of the diameter
 ratio $\alpha\equiv \sigma_2/\sigma_1$, both in 4D and 5D. As clearly seen
 in the
figure, the agreement is virtually perfect.
This agreement has been amply confirmed for other values of the concentration $x_{1}$ and the diameter ratio $\alpha$ \cite{Jose}.

We next consider the instability of a binary mixture ($N=2$) by looking at 
the
Helmholtz free energy  per unit volume, $a$, which is given by
\begin{equation}
\frac{a}{\rho k_{B}T}= -1+\sum_{i=1}^{2}x_{i}\ln \left(
\rho_{i}\lambda_{i}^d\right) +\int_{0}^{\eta }\frac{\upd\eta \prime}{\eta
\prime}\left[
Z_{\text{m}}(\eta \prime )-1\right] ,  \label{FEN}
\end{equation}
 where  $\lambda_{i}$ is the thermal de Broglie wavelength of species
$i$.
We locate the spinodals through the condition
$a_{11}a_{22}-a_{12}^{2}=0$,
with $a_{ij}\equiv \partial ^{2}{ a}/\partial \rho _{i}\partial \rho
_{j} $. Due to the spinodal instability, the mixture separates into two
phases of different composition. The coexistence conditions are
determined
through the equality of the pressure $p$ and the two
chemical
potentials $\mu _{1}$ and $\mu _{2}$ in both phases ($\mu _{i}=
\partial {a}/\partial \rho _{i}$), leading to
binodal curves. It should be noted that in our calculations we have
 considered the range $\eta _{1}+\eta
_{2}\leq \eta _{\text{cp}}$
($\eta_{\text{cp}}=\pi^2/16\simeq 0.617$ in 4D and
$\eta_{\text{cp}}=\pi^2\sqrt{2}/30\simeq 0.465$ in 5D \cite{LM90}) and
that
without
loss of generality we have set $k_{B}T=1$. Also, for future reference it
is
useful to recall that the freezing transition in the same systems occurs
at
a packing fraction $\eta _{\text{f}}\simeq 0.31$  and $\eta_{\text{f}}\simeq
0.20$,
respectively \cite{MT84}.

\begin{figure}
\twofigures[scale=1,clip=]{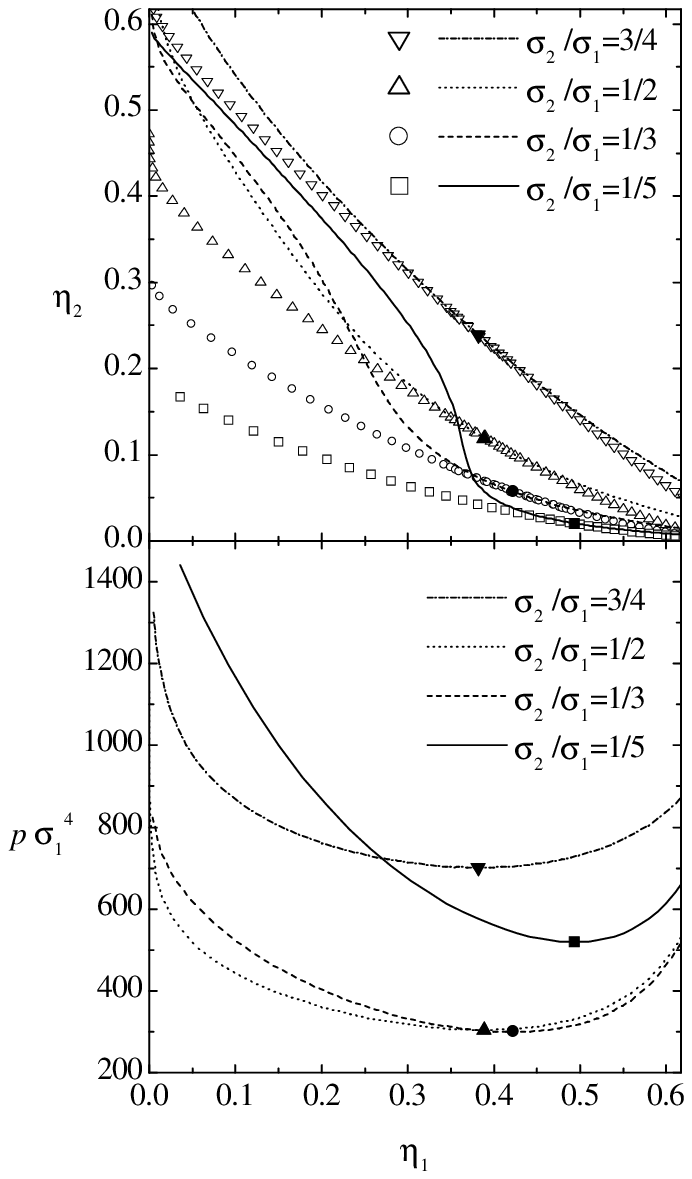}{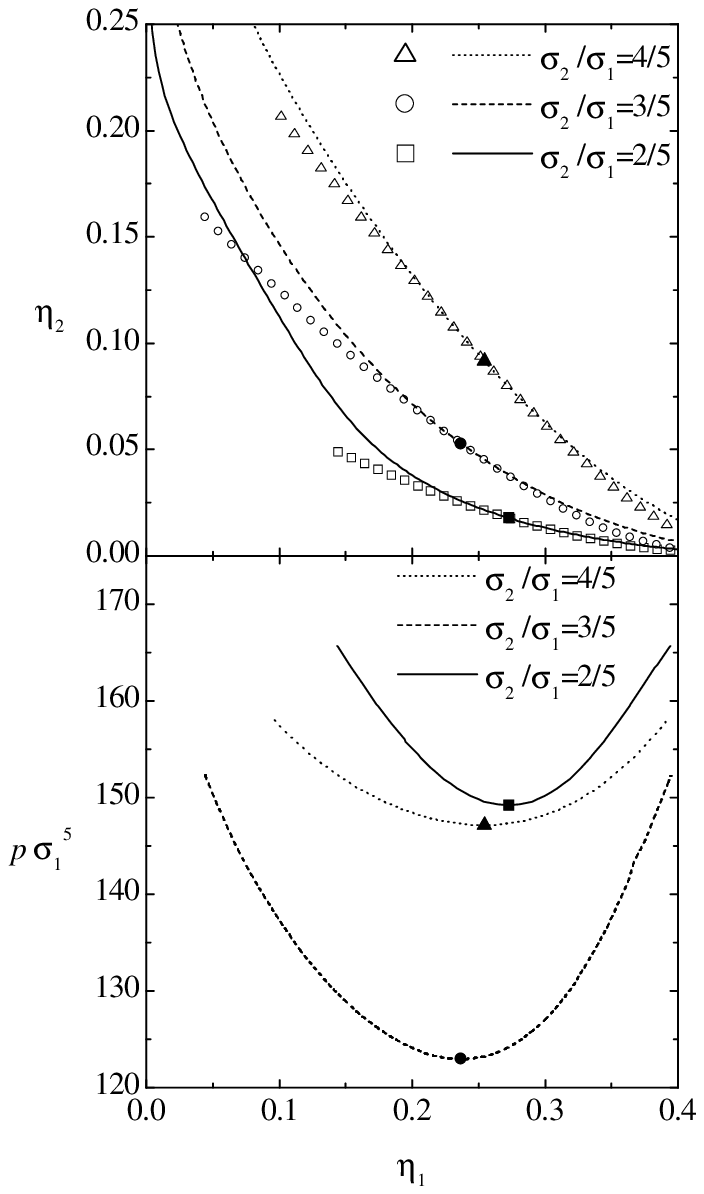}
\caption{Spinodal curves (upper panel: lines) and binodal
curves (upper panel: symbols; lower panel: lines) in a 4D system. The
filled symbols are the  critical consolute points.
\label{fig2}}
\caption{Same as Fig.\ \protect\ref{fig2}, but for a 5D system.
\label{fig3}}
\end{figure}

Figures \ref{fig2} and \ref{fig3} show the results for the spinodals and
binodals in the $\eta_{2}$ vs $\eta _{1}$ plane  and for the binodals in
the $p\sigma _{1}^{d}$ vs $\eta _{1}$ plane for
different values of $\alpha$
 and for $d=4$ and $d=5$, respectively. The location of the
critical point  tends to go down and to the right in the
$\eta_{2}$ vs $\eta _{1}$ plane as $\alpha $ decreases. Also, in
both dimensionalities and for a given $\alpha $,  the binodal and
spinodal curves become practically indistinguishable to the right
of the critical point. In contrast,  while the binodals with the
bigger $\alpha $ are always above those corresponding to a lower
diameter ratio, to the left of the critical point the different
spinodal curves cross each other  for $d=4$. The value of the
critical pressure $p_{\text{c}}$ (in units of $\sigma _{1}^{-d}$)
is not a monotonic function of $\alpha $; its minimum value lies
between $\alpha =1/3$ and $\alpha =1/2$ when $d=4$, and it is
around $\alpha =3/5$ for $d=5$. This non-monotonic behavior was
also observed  for hard spheres \cite{CB98b}.

It is conceivable that the demixing transition in binary mixtures of
hard hyperspheres in four and five dimensions described above may be
metastable with respect to a fluid-solid transition (cf.\ the values of 
$\eta _{\text{f}}$), as it also occurred with the case of hard spheres 
\cite{CB98b}. In fact the value of the pressure at the freezing transition
for the single component fluid is $p_{\text{f}}\sigma ^{d}\simeq 12.7$ ($d=3$), 
11.5 ($d=4$),  and 12.2 ($d=5$), i.e.\ $p_{\text{f}}\sigma ^{d}$ does not
change appreciably with the dimensionality but is clearly very small in
comparison with the critical pressures $p_{\text{c}}\sigma _{1}^{d}$
we obtain for the mixture; for instance, $p_{\text{c}}\sigma_1^{d}\simeq 
1000$ ($d=3$, $\alpha=3/10$)\cite{CB98b}, 300 ($d=4$, $\alpha=1/3$), and 123 
($d=5$, $\alpha=3/5$). However,
one should also bear in mind that, if the concentration $x_{1}$ of
the bigger spheres decreases, 
the value of the pressure at
which the solid-fluid transition in the mixture occurs in 3D is also considerably
increased with respect to $p_{\text{f}}$ (cf.\ Fig.\ 6 of Ref.\ \cite
{CB98b}). Thus, for concentrations $x_{1}\simeq 0.01$  corresponding to
the critical point of the fluid-fluid transition, the maximum pressure of
the fluid phase greatly exceeds $p_{\text{f}}$.  
If a
similar trend with composition also holds in 4D and 5D, and given that
the critical
pressures become smaller as the dimensionality $d$ is increased, it
is not clear whether the competition between the fluid-solid and the
fluid-fluid transitions in these dimensionalities will always be won by the
former. The point clearly deserves further investigation.

\begin{figure}
\twofigures[scale=1,clip=]{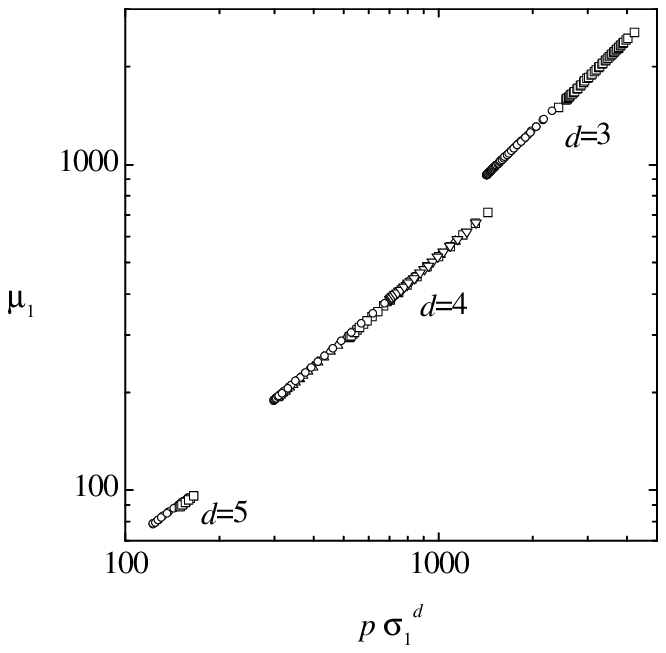}{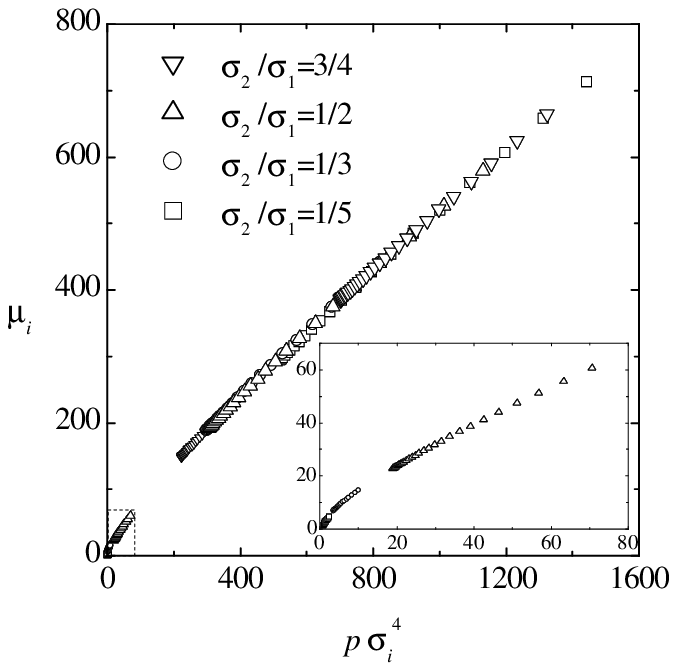}
\caption{Binodal
curves in the  $\mu_1$ vs $p\sigma_1^d$ plane for $d=3$ (circles: 
$\sigma_2/\sigma_1=1/5$, squares: $\sigma_2/\sigma_1=1/
10$), $d=4$ (symbols as in Fig.\ \protect\ref{fig2}), and $d=5$ (symbols as
in Fig.\ \protect\ref{fig3}).
Note that for each size ratio the binodal curve is restricted to
 $p\geq p_{\text{c}}$.
\label{fig4}}
\caption{{Binodal
curves in the  $\mu_i$ vs $p\sigma_i^4$ plane for a 4D system.
The case $i=1$ corresponds to $\mu_i\gtrsim 200$ and is represented by
the large symbols, while the case $i=2$ corresponds to $\mu_i\lesssim
200$ and is represented by the small symbols.
The inset is a 
magnification of the box enclosed by the dashed lines and shows $\mu_2$ vs 
$p\sigma_2^4$ for $\sigma_2/\sigma_1=1/2$, 1/3 and 1/5.}
\label{fig5}}
\end{figure}

An interesting feature arising in our results, and which to our
knowledge
has not been discussed up to now, must be mentioned. There is a
remarkable
similarity between the binodal curves represented in the 
$p\sigma_{i}^{d}$--$\eta _{1}$ and in the $\mu _{i}$--$\eta _{1}$ planes (not
shown). By
eliminating $\eta _{1}$ as if it were a parameter, 
 one can
represent the binodal curves in a $\mu _{i}$ vs
$p\sigma_{i}^{d}$ plane. This is shown in Figs.\ \ref{fig4} {and 
\ref{fig5}}, where we have chosen the origin
of
the chemical potentials such as to make $\lambda_i=\sigma_i$.
The
binodals  in the $\mu _{1}$--$p\sigma _{1}^{d}$ plane collapse into a
single universal curve (which is in fact almost a straight line) for
each dimensionality ($d=3$, $d=4$, and $d=5$).  We have obtained the data 
for $d=3$
by  using the EOS due to Coussaert and Baus\cite{CB98b}. For each 
dimensionality, the curve $\mu _{1}=F_{1}\left( p\sigma _{1}^{d}\right) $ is 
seen to be
independent of the asymmetry in sizes as the contributions from each 
$\alpha $ to a portion of the curve overlap. Something similar occurs
with the curve $\mu _{2}=F_{2}\left( p\sigma _{2}^{d}\right)$  and the 
calculations indicate that $F_{1}=F_{2}$  in the
sense that if plotted on the same plane, both curves seem to probe different
regions of the plane but otherwise overlap reasonably well. {This is 
illustrated by Fig.\ \ref{fig5} for the case $d=4$, where  
the points of $\mu _{2}$ corresponding to $\alpha =3/4$
connect with those of $\mu _{1}$ for $\alpha =1/3$ and 
$\alpha =1/2$}. These properties could reveal 
the existence
of a certain {\em geometrical\/} regularity that might characterize the
entropy driven transitions. One could then state that all
binary hard (hyper)sphere mixtures belong to the same {\em universality class\/}
since, for fixed $d$, there are some   properties such as the binodal curves 
in
which the asymmetry is to a certain extent irrelevant. We also find an upper
quasi-universality: for the range of common values, the universal curve in 
3D  lies close but somewhat above the universal curve in 4D,
and in turn this latter lies above the one in 5D (note, however, that the origin of $\mu_i$ is actually a matter of choice). All three curves
approach straight lines of slopes 0.44 ($d=3$), $0.45$ ($d=4$), and $0.43$ ($d=5$) in the region of their
higher values.

The  results  presented in this Letter
confirm that the geometrical effects of
osmotic depletion become more important as the dimensionality increases
in
much the same way that in 3D they are more important for parallel hard
cubes than for hard spheres. It could be argued that the results we
have
presented so far (except for the universality, which is
also found with a completely independent EOS in 3D) strongly depend on our 
prescription (\ref{4.1}) for the EOS of the
mixture and on the use of eq.~(\ref{4}). To address the first point, we have 
carried out
calculations
using
the van der Waals one-fluid theory with the Luban-Michels EOS in 4D and
found no spinodal instability for $\eta _{1}+\eta _{2}\leq \eta
_{\text{cp}}$. Concerning the second point, we have also performed
calculations using Eq.\
(\ref{4.1}) for $Z_{\text{m}}(\eta )$ with the EOS proposed by Baus and Colot
\cite{CB86} for the single component fluid. In this
instance
we also find spinodal curves which do not coincide but are qualitatively
similar with the ones obtained using eq.~(\ref{4}).  The sensitivity of
the
spinodal
curves with respect to the EOS was also noted  in
the
case of hard spheres\cite{CB98b}.
Finally, we want to mention that
our approach opens up the possibility of examining whether there is a
critical dimension above which the fluid-fluid transition in hard
hypersphere mixtures becomes stable, including the role that the asymmetry
plays in it. Also we hope that the universal feature that we have described
will stimulate other studies to clarify its physical origin and further
implications.

\acknowledgments

Two of us (A.S. and S.B.Y.) acknowledge partial financial support from
DGES (Spain) through Grant No.\ PB97-1501 and from the Junta de
Extremadura
(Fondo Social Europeo) through Grant No.\ IPR99C031. The work of M.L.H.
has
been partially supported by DGAPA-UNAM under project IN-117798.

\end{document}